\documentclass[twoside,preprintnumbers,pacs,twocolumn,nofootinbib]{revtex4}
\usepackage{amsmath}
\usepackage{epsfig}
\usepackage{graphicx}
\usepackage{dcolumn}
\usepackage{bm}
\usepackage{braket}
\usepackage{amsthm}
\usepackage{pslatex}

\newcommand{\occ}{\overline{c}}

\begin{document}

\title{Gribov no-pole condition, Zwanziger horizon
function, Kugo-Ojima confinement criterion, boundary conditions, BRST
breaking and all that}
\author{D. Dudal$^a$}
\author{S.P. Sorella$^b$}
\author{N. Vandersickel$^a$}
\author{H. Verschelde$^a$}
\email{david.dudal@ugent.be, sorella@uerj.br, nele.vandersickel@ugent.be,
henri.verschelde@ugent.be}
\affiliation{\vskip 0.2cm $^a$ Ghent University, Department of Mathematical Physics and
Astronomy, \\
Krijgslaan 281-S9, B-9000 Gent, Belgium\vskip 0.1cm$^b$ Departamento de F%
\'{\i}sica Te\'{o}rica, Instituto de F\'{\i}sica,UERJ - Universidade do
Estado do Rio de Janeiro,\\
Rua S\~{a}o Francisco Xavier 524, 20550-013 Maracan\~{a}, Rio de Janeiro,
Brasil}

\begin{abstract}
\noindent We aim to offer a kind of unifying view on two popular topics in the studies of nonperturbative aspects of  Yang-Mills
theories in the Landau gauge: the so-called Gribov-Zwanziger approach and
the Kugo-Ojima confinement criterion. Borrowing results from statistical
thermodynamics, we show that imposing the Kugo-Ojima confinement criterion
as a boundary condition leads to a modified yet renormalizable partition
function. We verify that  the resulting partition function is
equivalent with the one  obtained by Gribov and Zwanziger, which
restricts the  domain of integration in the path integral within
the first Gribov horizon.  The construction of an action
implementing a boundary condition allows one to discuss the symmetries of
the system in the presence of the boundary. In particular, the conventional
BRST symmetry is softly broken.
\end{abstract}

\maketitle

\setcounter{page}{1}

\section{Introduction}
The Gribov-Zwanziger (GZ) approach focuses on the issue of gauge copies in
the Landau gauge. Gribov signalled in his seminal work \cite{Gribov:1977wm}
that the Landau gauge condition, $\partial _{\mu }A_{\mu }=0$ is ambiguous:
there exist gauge equivalent configurations $A_{\mu }^{\prime }$ which also
obey $\partial _{\mu }A_{\mu }^{\prime }=0$.  Examples of gauge
copies are provided by the zero modes of the Faddeev-Popov (FP) operator, which
enters the quantization  formula of Yang-Mills theories. Indeed,
given an infinitesimal gauge transformation connecting $A_{\mu }$ with $%
A_{\mu }^{\prime }$, i.e. $A_{\mu }^{a^{\prime }}=A_{\mu }^{a}-D_{\mu
}^{ab}\omega ^{b}$, it is clear that $\partial _{\mu }A_{\mu }^{\prime
}=\partial _{\mu }A_{\mu }=0$ is fulfilled when $M^{ab}\omega ^{b}=0$, with $%
M^{ab}=-\partial _{\mu }D_{\mu }^{ab}=-\partial _{\mu }\left( \partial _{\mu
}\delta ^{ab}+gf^{acb}A_{\mu }^{c}\right) $  being the
FP operator. We recall that the FP action
in the Landau gauge for a $d$-dimensional Euclidean gauge theory, with $d\leq 4$, reads
\begin{equation}
S_{\mathrm{YM+gf}}=S_{\mathrm{YM}}+\int {d^{{d}}}x\;\left( b^{a}\partial
_{\mu }A_{\mu }^{a}+\overline{c}^{a}\partial _{\mu }D_{\mu
}^{ab}c^{b}\right)  \label{action} \,,
\end{equation}%
with $S_{\mathrm{YM}}=\frac{1}{4}\int {d^{d}}xF_{\mu \nu }^{a}F_{\mu \nu }^{a}
$ the classical Yang-Mills action.  Expression \eqref{action}
enjoys the well-known BRST symmetry,  generated by the nilpotent
operator $s$, $s^{2}=0$, i.e.
\begin{equation}\label{BRST}
sA_{\mu }^{a}=-D_{\mu }^{ab}c^{b}\;,sc^{a}=\frac{1}{2}gf^{abc}c^{b}c^{c}\;,s%
\overline{c}^{a}=b^{a}\;,sb^{a}=0\,.
\end{equation}
For the partition function, we can write
\begin{equation}
Z=\int d\mu _{\mathrm{FP}}=\int d\Phi e^{-S_{\mathrm{YM+gf}}}=\int dA\det M^{ab}\delta
(\partial A)e^{-S_{\mathrm{YM}}}\,.  \label{part1}
\end{equation}%
We introduced the notational shorthand $\Phi $ denoting all the fields
present in the action, with $d\mu _{\mathrm{FP}}$ the usual FP measure.

Gribov proposed to restrict the domain of integration to the subspace $\Omega $, where the Hermitian operator $M^{ab}$ is positive definite. More
precisely, we define the Gribov region as $\Omega \equiv \{A_{\mu }^{a},\;\partial _{\mu }A_{\mu }^{a}=0,\;M^{ab}>0\}$. We recognize that configurations $A_{\mu }^{a}\in \Omega $ are relative
minima of the functional $\int d{^{d}}x(A_{\mu }^{u})^{2}$, $u\in SU(N)$.
The boundary, $\partial \Omega $, of $\Omega $ is called the (first) Gribov
horizon. It was shown with increasing rigor that $\Omega $ is
convex,  bounded in all directions in field space, and that every
gauge field has at least one gauge equivalent representant in $\Omega $ (see
\cite{Zwanziger:1989mf,Zwanziger:1992qr} and references therein). The inverse  of the FP operator, or equivalently
the ghost propagator with external gauge field, $G^{ab}(k,A)$, can be used
to implement the restriction to $\Omega $, as done semiclassically by
Gribov.  Following \cite{Gribov:1977wm}, we can write
\begin{equation}
G^{ab}(k,A)=\frac{\delta ^{ab}}{k^{2}}\frac{1}{1+\sigma (k,A)}=(M^{-1})^{ab}(k,A)\,.  \label{GR1}
\end{equation}%
At lowest order, it can be shown that $1+\sigma (k,A)$ is a decreasing
function of $k$  \cite{Gribov:1977wm}, hence one can impose
\begin{equation}
1+\sigma (0,A)\geq 0\,.  \label{np}
\end{equation}%
Condition \eqref{np}, known as the Gribov no-pole condition,
implies that the ghost propagator $G^{ab}(k,A)$ has no poles at finite
nonvanishing $k$. Moreover, positivity of $G^{ab}(k,A)$ ensures that the
Gribov horizon $\partial \Omega $ is not crossed.  As done by
Gribov \cite{Gribov:1977wm}, the no-pole condition can be embodied into the
partition function using a $\delta $-function\footnote{
Condition \eqref{np} can be implemented by inserting a step function factor \mbox{$\theta (1+\sigma (k,A))$}. However, in the
thermodynamic limit, the $\theta $-function can be can be replaced by a $%
\delta $-function, see \cite{Gribov:1977wm,boek,Zwanziger:1989mf}.},
\begin{equation}\label{zprime}
Z^{\,\prime }=\int d\Phi \delta (1+\sigma (0,A))e^{-S_{\mathrm{YM+gf}}}\,.
\end{equation}
Later on, Zwanziger  \cite{Zwanziger:1989mf} was able to implement the
no-pole condition to all orders.  Relying on the equivalence
between the microcanonical and the canonical Boltzmann ensemble (see also
Section II),  he was able to show that the partition function %
\eqref{zprime} has to be replaced by
\begin{equation}
Z^{\,\prime \prime }=\int dA\delta (\partial _{\mu }A_{\mu })\det M^{ab}e^{-S_{%
\mathrm{YM}}+\gamma ^{4}\int d{^{d}}xh(x)}\,,  \label{part3d}
\end{equation}%
 where the Zwanziger horizon function reads
\begin{equation}
\int d{^{d}}xh(x,A)=g^{2}\int d{^{d}}xd^{d}yf^{abc}A_{\mu
}^{b}(x)(M^{-1})^{ad}(x,y)f^{dec}A_{\mu }^{e}(y)  \label{part3a}
\end{equation}%
The mass parameter $\gamma $ is
determined by a gap equation, commonly called the horizon condition
\begin{equation}
\braket{h(x)}=d(N^{2}-1)\,,  \label{part3}
\end{equation}%
where $\braket{h}$ is calculated with $Z^{\,\prime \prime }$,
i.e. with the measure $d\mu _{\mathrm{FP}}e^{\gamma ^{4}\int d{^{d}}xh(x)}$.
The  factor $d(N^{2}-1)$ in the r.h.s. was obtained \cite{Zwanziger:1989mf}  by determining the lowest eigenvalue of the
FP operator. Working out the condition \eqref{part3} at lowest
order reproduces the Gribov result \cite{Gribov:1977wm}. The action
corresponding to the partition function \eqref{part3d} contains the \emph{%
nonlocal} horizon term \eqref{part3a}. To arrive at a workable quantum model, it was shown
\cite{Zwanziger:1992qr} that \eqref{part3d} can be put in an equivalent
\emph{local} form by introducing a set of complex conjugate commuting
variables, $(\varphi _{\mu }^{ac},\overline{\varphi }_{\mu }^{ac})$, and
anticommuting ones, $(\omega _{\mu }^{ac},\overline{\omega }_{\mu }^{ac})$,
so that we finally obtain the Gribov-Zwanziger action,
\begin{eqnarray}
S_{\mathrm{GZ}} &=&S_{\mathrm{YM+gf}}+\int d{^{{d}}}x\left( \overline{%
\varphi }_{\mu }^{ac}\partial _{\nu }D_{\nu }^{ab}\varphi _{\mu }^{ac}-%
\overline{\omega }_{\mu }^{ac}\partial _{\nu }D_{\nu }^{ab}\omega _{\mu
}^{ac}\right.   \notag  \label{snul} \\
&&\left. -g\left( \partial _{\nu }\overline{\omega }_{\mu }^{ac}\right)
f^{abm}\left( D_{\nu }c\right) ^{b}\varphi _{\mu }^{mc}\right.   \notag \\
&&\left. -\gamma ^{2}gf^{abc}A_{\mu }^{a}\left( \varphi _{\mu }^{bc}+%
\overline{\varphi }_{\mu }^{bc}\right) +d\left( N^{2}-1\right) \gamma
^{4}\right) \,.
\end{eqnarray}
The horizon condition \eqref{part3} is translated as $\frac{\partial \Gamma
}{\partial \gamma }=0$, with $\Gamma (\gamma )$ the effective action,
defined as $e^{-\Gamma }=\int d\Phi e^{-S_{\mathrm{GZ}}}$. This can be
easily checked, given that we take $\gamma \neq 0$.  The mass
parameter $\gamma $  turns out to be proportional to $\Lambda _{%
\overline{\mbox{\tiny{MS}}}}$, and as such it can give rise to
nonperturbative corrections. This is not unexpected, as the restriction to the region $\Omega $ is a highly nontrivial operation, which
goes beyond perturbation theory. At the perturbative level, the ghost
propagator stays positive.  We are thus far from the horizon and
nothing happens. It is only at lower momenta, where normal perturbation
theory starts to fail, that the fields begin to feel the
restriction to $\Omega $. Having brought the action in standard local form, we have all the usual
concepts and machinery of local quantum field theory to our disposal. A
first important property of \eqref{snul} is its renormalizability to all
orders of perturbation theory. Hence, the restriction to $\Omega$
makes perfect sense at
the quantum level, and finite results are found, consistent with the
renormalization group \cite{Zwanziger:1992qr,Dudal:2005na}. We stress here
that the action \eqref{snul}, with the horizon condition \eqref{part3}
implemented, is nothing else than the correct extension to all orders of the
usual Yang-Mills action, supplemented with the Landau gauge fixing, in the
presence of a nontrivial boundary condition, being the no-pole condition %
\eqref{np}. In this fashion, it is assured that we have taken care of a certain amount of gauge copies, including those related to the
zero modes of $M^{ab}$. Notice that this does not mean that the Gribov issue has been completely solved. It is known that $\Omega $
still contains copies, related to the fact that $\int d{^{d}}x(A_{\mu
}^{u})^{2}$ can have many relative minima starting from the same $A_{\mu }$.
A further restriction is needed, keeping only gauge configurations that are
absolute minima of $\int {d^{d}}x(A_{\mu }^{u})^{2}$; the latter define the
fundamental modular region (FMR) $\Lambda $.

Evidently, the extra fields,$(\overline{\varphi}_\mu^{ac}$, $%
\varphi_\mu^{ac} $, $\overline{\omega}_\mu^{ac}$,$\omega_\mu^{ac}$), can
influence the dynamics of the theory in a nontrivial fashion \cite%
{Dudal:2007cw}.  These fields arise as a consequence of the
presence of the Gribov horizon. As such, they can give rise to
additional nonperturbative effects. For example, in \cite{Dudal:2008sp}, we have provided evidence of the existence of a dimension 2 condensate, $\braket{\overline{\varphi}_{\mu}^{ac}\varphi_{\mu}^{ac}-\overline{\omega}_{%
\mu}^{ac}\omega_{\mu}^{ac}}$, in $d=4$. A posteriori, this is not that
surprising, given that the restriction to $\Omega$ introduces the mass scale $\gamma$ into the theory, and that the horizon condition \eqref{part3} can be
reexpressed at the local level as $\braket{gf^{abc}A_\mu^a(\varphi_{%
\mu}^{ac}+\overline{\varphi}_\mu^{ac})}=-2\gamma^2d(N^2-1)$, i.e. a
dimension 2 condensate for $d=4$. Nontrivial condensates are an important source of
nonperturbative effects in gauge theories, hence the general interest in
their study. In particular, dimension 2 condensates attracted a lot of
attention in recent years, see e.g. \cite{Dudal:2005na,Chernodub:2008kf} and
references therein. In the current case, the operator $(\overline{\varphi}%
_{\mu}^{ac}\varphi_{\mu}^{ac}-\overline{\omega}_{\mu}^{ac}\omega_{\mu}^{ac})$
can be added to the theory  in a way that preserves
renormalizability \cite{Dudal:2008sp}, which is already a remarkable
feature, indicative of its possible relevance. We studied the effects
of  this condensate using variational perturbation theory, and
found that the gluon propagator\footnote{ The Landau gluon propagator can be parametrized in
terms of the form factor $D(k^2)$ as $\braket{
A^a_{\mu}(k)A^b_\nu(-k)}= \delta^{ab}(\delta_{\mu\nu}-\frac{k_\mu k_\nu%
}{k^2})D(k^2)$.} does not vanish at zero momentum ($D(0)\neq0$%
), that the ghost propagator behaves like $\sim%
\frac{1}{k^2}$ at small momenta, and that there is a violation of positivity
in the gluon propagator \cite{Dudal:2008sp}. Any of these findings is in
good agreement with \emph{all} most recent lattice data, obtained at
previously unseen large volumes \cite{Cucchieri:2007md,Maas:2008ri}. Also
certain results based on Schwinger-Dyson (SD) and/or Functional Renormalization Group (FRG)
equations are consistent with  these data, see e.g.
\cite{Aguilar:2008xm,Fischer:2008uz}. Without taking into account the
effects related to $\braket{\overline{\varphi}_{\mu}^{ac}\varphi_{\mu}^{ac}-%
\overline{\omega}_{\mu}^{ac}\omega_{\mu}^{ac}}$, the GZ action \eqref{snul}
also leads to the positivity violation of the gluon propagator,
however with $D(0)=0$, and an infrared enhanced ghost. These latter two results are no longer supported by lattice data. Hence, it seems
crucial to take into account additional nonperturbative effects related to
the restriction  to the region $\Omega$ (i.e. the boundary
condition) to allow for consistency between the analytical GZ results and
most recent lattice predictions. The interpretation of the analytical \cite{Dudal:2007cw,Dudal:2008sp} and the lattice
results of \cite{Cucchieri:2007md} was challenged in papers like \cite{Maas:2008ri,Fischer:2008uz,Sternbeck:2008mv}. It was argued that the ghost propagator
must be infrared enhanced to ensure confinement, whereby only colorless
states are physical. These statements are based on
the Kugo-Ojima (KO) analysis of gauge theories \cite{Kugo:1979gm,Kugo:1995km}%
. This analysis  relies on the operator formalism, and it has been shown that, given a globally well-defined BRST charge $%
Q_{B}$, the color charge $Q^{a}$ is a BRST exact variation, $Q^{a}=Q_{B}(\ldots )$, if the gluon propagator contains no massless poles.
The color charge $Q^{a}$ is then well-defined only if the KO
confinement criterion holds
\begin{equation}
u(0)=-1\,,  \label{KO0}
\end{equation}%
with $u(k^{2})$ defined through the following Green function
\begin{equation}\label{KO1}
\int d{^{d}}\!xe^{ikx}\Braket{D_\mu^{ad}c^d(x) D^{\nu\;be}\occ^e(0)}_{\mathrm{%
FP}}  =\delta ^{ab}\left( P_{\mu\nu}(k)u(k^{2})-\frac{k_{\mu }k_{\nu }}{k^{2}}\right)
\end{equation}
in Minkowski space. $\braket{\ldots}_{\mathrm{FP}}$ stands for the expectation value taken with the
FP action \eqref{part1}, while $P_{\mu\nu}(k)=g_{\mu \nu }-k_{\mu }k_{\nu }/k^{2}$ for the transverse projector. Using the nilpotent BRST charge $Q_{B}$,
one can invoke its cohomology to define the physical subspace, and by means
of $Q^{a}=Q_{B}(\ldots )$, conclude that physical states cannot carry color. A few comments are in order. First of all,  in the KO framework
\cite{Kugo:1979gm,Kugo:1995km}, the existence of a globally well-defined
BRST charge is \emph{assumed}.  Thus, the issue of the
(non)existence of a nonperturbatively valid BRST symmetry  is not
explicitly faced. Secondly, the link between the BRST charge $Q_B$ and
global color charge $Q^a$ is made using the action \eqref{action}, i.e. by
employing the usual FP gauge fixed action. As such, the Gribov
problem is  simply not addressed. It is worth noticing
that Kugo and Ojima did not impose the criterion \eqref{KO0}, but they
derived it as a condition to be checked/calculated.  Though,
nowadays, in functional formalisms as in \cite{Fischer:2008uz}, the
criterion is used as \emph{input}. Kugo showed in \cite{Kugo:1995km} that, in the Landau gauge, one can rewrite the ghost
propagator
\begin{equation}  \label{ghpro}
G^{ab}(k)=\frac{\delta^{ab}}{k^2}\frac{1}{1+u(k^2)+k^2v(k^2)}\,,
\end{equation}
meaning that the criterion \eqref{KO0} is
equivalent to an infrared enhanced ghost. The ghost enhancement is then imposed as a boundary condition in
order to favor the so-called scaling type solution of the SD and/or FRG
equations \cite{Fischer:2008uz}. Let us already draw attention to the close similarity existing between the no-pole
condition \eqref{np} and the criterion \eqref{KO0}. Imposing\footnote{$\sigma(0)$ is related to $\sigma(0,A)$ by making the gauge field
dynamical and performing the corresponding path integration.} $\sigma(0)=-1$
exactly corresponds to $u(0)=-1$.

\section{$u(0)=-1$ as a boundary condition}
We want to show that  the constraint $u(0)=-1$  can
be implemented directly into the theory, by appropriately modifying the
measure one starts from. We shall see that the  resulting action
will be exactly the same as the GZ action. This has  several
interesting consequences which we will discuss in Section III. We shall
first give an overview of some results from thermodynamics we intend to
employ.

\subsection{Microcanonical ensemble and equivalence with the canonical
Boltzmann ensemble in the thermodynamic limit}
We consider a discrete system, whose Hamiltonian is $H(q,p)$, with $3N$ degrees of freedom. The averages in the microcanonical ensemble
are constructed out of
\begin{equation*}
\Sigma (E)=\int_{H=E}d\mu \ =\int d\mu \;\delta (E-H)\,,
\end{equation*}
where $d\mu=d^{3N}qd^{3N}p$ represents the classical phase space and $E$ stands for the constant energy of the system. Averages in the microcanonical ensemble are defined by $
\braket{O}_{\mathrm{Micr}}=\frac{\int_{H=E}d\mu \ O}{%
\int_{H=E}d\mu }$. In order to establish the equivalence between the microcanonical and the
(Boltzmann) canonical ensemble we rewrite the quantity $\Sigma (E)$ in the
following form
\begin{align}
\Sigma (E)& =\int d\mu \ \delta (E-H)=\int d\mu \int_{-i\infty +\varepsilon
}^{i\infty +\varepsilon }\frac{d\beta }{2\pi i}e^{\beta (E-H)}  \notag \\
& =\int \frac{d\beta }{2\pi i}f(\beta )\ =\int \frac{d\beta }{2\pi i}%
e^{-\omega (\beta )}\,,  \\
f(\beta )&=\int d\mu ~e^{\left( \beta (E-H)\right) }\;,\qquad \omega (\beta
)=-\log f(\beta )\,.  \label{dd5}
\end{align}
It can be shown  that, in the thermodynamic limit, $%
N,V\rightarrow \infty $, with $N/V$ fixed, the saddle point approximation
becomes exact. We refer to \cite{boek} for an overview of the proof. So,
\begin{equation}
\Sigma (E)=\frac{1}{2\pi i}f(\beta ^{\star })\;,\text{with}\;\omega ^{\prime
}(\beta ^{\star })=\frac{f^{\prime }(\beta ^{\star })}{f(\beta ^{\star })}%
~=~0\,.  \label{d8}
\end{equation}%
From eq.$\left( \ref{d8}\right) $ it follows that
\begin{equation}
E=\braket{H}_{\mathrm{Boltz}}=\frac{\int d\mu ~He^{-\beta ^{\star }H}}{\int d\mu ~e^{-\beta ^{\star }H}}%
\,.  \label{d10}
\end{equation}%
This is the gap equation determining the critical parameter $\beta ^{\star }$. Analogously, it can also be shown that \cite{boek} $
\braket{O}_{\mathrm{Micr}}=\braket{O}_{\mathrm{Boltz}}=\frac{\int d\mu ~O\ e^{-\beta ^{\star }H}}{\int d\mu
~e^{-\beta ^{\star }H}}$ for the average of any quantity $O(q,p)$.

\subsection{ Imposing the KO criterion yields the
GZ framework}
Starting from \eqref{KO1} and performing Lorentz and color contractions and
taking the $p\rightarrow 0$ limit, we can write
\begin{eqnarray}
&&-(VT)^{-1}\int d{^{d}}y\int d{^{d}}x\braket{D_\mu^{ad}(x)D_\mu^{ae}(y)
(M^{-1})^{de}(x,y)}_{\mathrm{FP}}
\notag  \label{KO2} \\
&=&(N^{2}-1)((d-1)u(0)-1)\,,
\end{eqnarray}%
after passing to Euclidean space, as in any functional or lattice approach. $VT$ denotes the spacetime volume. The identification between $\braket{\ldots c^d(x)\occ^e(y)}_{\mathrm{FP}}$
and $\braket{\ldots (M^{-1})^{de}(x,y) }_{\mathrm{FP}}$ can be easily proven
using the path integral \eqref{part1}. After discarding terms which are
total derivatives, one easily sees that the quantity in the l.h.s. of %
\eqref{KO2} is, up to the sign,  the Zwanziger horizon
function $h(x)$. More precisely, we have
\begin{eqnarray}
&&\hspace{-0.8cm}\eqref{KO2}=\int \frac{d{^{d}}y}{VT}d{^{d}}x%
\braket{gf^{akd}A_\mu^k(x) (M^{-1})^{ed}(x,y) gf^{ame}A_\mu^m(y)}_{\mathrm{FP}}
\notag \\
&=&-(VT)^{-1}\int {d^{{d}}}x\braket{h(x)}_{\mathrm{FP}}=-\braket{h}_{\mathrm{%
FP}}\,.  \label{d16}
\end{eqnarray}%
We observe that the KO condition cannot be realized
with the standard FP measure $d\mu _{\mathrm{FP}}$, otherwise we would have
\begin{equation}
\braket{h(x)}_{\mathrm{FP}}=d(N^{2}-1)\,,\label{ex1}
\end{equation}%
which would contradict Zwanziger's result (\ref{part3}), obtained by
restricting the path integral to the Gribov region $\Omega $. We now implement the KO criterion $u(0)=-1$ as a boundary
condition, amounting to start from the modified measure
\begin{equation}
d\mu _{\mathrm{FP}}\;\to\; d\mu^{\,\prime}\equiv d\mu _{\mathrm{FP}}\ \delta \left(
VTd(N^{2}-1)-\int d^{d}xh(x)\right) \,,  \label{ex2}
\end{equation}%
which clearly implements $\braket{h(x)} =d(N^{2}-1)$, or equivalently $u(0)=-1$. We are thus led to consider the partition function
\begin{eqnarray}
&&\hspace{-7mm}\int d\mu^{\,\prime}~=~\int d\mu _{\mathrm{FP}}\ \delta \left( VTd(N^{2}-1)-\int
d^{d}xh(x)\right)\label{d20}\notag \\
&&\hspace{-5mm} = \int dA\ \delta (\partial A)\ \det \mathcal{M}\ e^{-S_{%
\mathrm{YM}}}\ \delta \left( VTd(N^{2}-1)-\int d^{d}xh(x)\right) \notag
 \\
&=&\int d\Phi \delta \left( VTd(N^{2}-1)-\int d^{d}xh(x)\right) e^{-S_{%
\mathrm{YM+gf}}}\,.
\end{eqnarray}%
Expression $\left( \ref{d20}\right) $ defines a microcanonical ensemble.
Since we are working in a continuum field theory, we are working in
the thermodynamic limit, hence we have an equivalence with a Boltzmann
canonical ensemble as outlined in the previous section. Using analogous
arguments as there, we arrive at
\begin{equation}
\int d\mu^{\,\prime}
=\int d\mu _{\mathrm{FP}}\ e^{\gamma ^{4}\int d{^{{d}}}x\,h(x)}\equiv\int d\mu _{\mathrm{FP}}\ e^{-S_\mathrm{H}}\,,
\label{d21}
\end{equation}%
where the mass parameter $\gamma $ follows from the gap equation
\begin{equation}
d\left( N^{2}-1\right) =\braket{h(x)}_{\mathrm{Boltz}}=%
\frac{\int d\mu _{\mathrm{FP}}\ e^{-S_{\mathrm{H}}}\ h(x)}{\int d\mu _{%
\mathrm{FP}}\ e^{-S_{\mathrm{H}}}}\,,  \label{d22}
\end{equation}%
which is the analogue of \eqref{d10}. We conclude that we can consistently encode the boundary condition \eqref{KO0} at the level of the
action, which turns out to be identical to the GZ action, eq.\eqref{part3d}. Of course, we can localize it into the form %
\eqref{snul}, with corresponding local formulation of the gap equation.

\section{Discussion}
Naively,  one might already expect that the introduction of a
nontrivial boundary condition can seriously influence the dynamics of the
theory. One of our main points is to stress that one should introduce the
boundary condition into the theory from the beginning, to fully grasp all its nontrivial aspects. Having at our disposal an action
automatically implementing the boundary  condition, we can study
an important aspect: the symmetries of the theory in the presence of the
boundary. In principle, imposing a boundary could jeopardize certain
symmetries of the original action. We have already shown in
\cite{Dudal:2008sp} that placing a boundary in  field space at
the first Gribov horizon breaks the conventional BRST
symmetry \eqref{BRST}. The practical implementation of the horizon by
means of the GZ formulation confirms this, as $sS_{\mathrm{GZ}}=g\gamma
^{2}\int {d^{d}}xf^{abc}\left( A_{\mu }^{a}\omega _{\mu }^{bc}-\left( D_{\mu
}^{am}c^{m}\right) \left( \overline{\varphi }_{\mu }^{bc}+\varphi _{\mu
}^{bc}\right) \right) \neq 0$. We notice that the BRST generator \eqref{BRST}
has a natural extension to the extra fields $(\overline{\varphi }_{\mu
}^{ac},\varphi _{\mu }^{ac},\overline{\omega }_{\mu }^{ac},\omega _{\mu
}^{ac})$, given by
\begin{equation}
s\varphi _{\mu }^{ac}=\omega _{\mu }^{ac}\;,s\omega _{\mu }^{ac}=0\;,s%
\overline{\omega }_{\mu }^{ac}=\overline{\varphi }_{\mu }^{ac}\;,s\overline{%
\varphi }_{\mu }^{ac}=0\,,  \label{BRSTbis}
\end{equation}%
forming 2 pairs of BRST doublets. In the absence of the GZ restriction or
equivalently  of the KO criterion, i.e. when $\gamma \equiv 0$,
we are then assured  that these fields are trivial in the BRST
cohomology, thus completely  decoupling from the physical
subspace \cite{Piguet:1995er}. Let us come to another important statement.
If Gribov copies are taken into account \`{a} la GZ, which is equivalent to
imposing the KO criterion as we have verified in the previous section, the
precise meaning of the KO confinement criterion becomes unclear. Since the
BRST symmetry is broken, one can no longer simply use it to define the
physical subspace. It is sometimes mentioned in the literature
that there might be a nonperturbative, globally well-defined BRST charge $%
Q_{B}^{\prime }$, and it is this $Q_{B}^{\prime }$ KO is referring to \cite%
{Fischer:2008uz,Sternbeck:2008mv}. We cannot exclude this possibility, but
this is a highly nontrivial statement and, obviously, it asks for a proof. At present, we are unaware of any such proof. Even if the charge
$Q_{B}^{\prime }$ would be known, the KO analysis would need to be reworked
from the start, as it explicitly relies on the FP action \eqref{action} and conventional BRST symmetry \eqref{BRST}. Simply stating
that $Q_{B}^{\prime }$ must exist  in order to define the
physical subspace analogously to  what is done at perturbative
level, does, in our opinion, not solve the problem. Also, the relation
between a new BRST charge $Q_{B}^{\prime }$ and the global color charge
would need to be reestablished, if any relation exists to begin with. One
can speculate that it might be possible to modify $s$ into $s_{\gamma }$,
such that $\lim_{\gamma \rightarrow 0}s_{\gamma }=s$ and $s_{\gamma }S_{%
\mathrm{GZ}}=0$. However, such a possibility can be easily disproved.
Indeed, as $\gamma $ has mass dimension 1, and by keeping in mind that the
BRST generator $s$ does not affect the dimension of the fields\footnote{%
The usual canonical dimensions are assigned to the fields \cite%
{Piguet:1995er}.}, it is impossible to introduce extra $\gamma $-dependent
terms in the BRST transformation of the fields while preserving locality,
Lorentz covariance and global $SU(N)$ structure. Let us briefly return to
the functional SD (FRG) approaches. Now that a renormalizable action has
been constructed, which implements the desired boundary condition
explicitly, one can write down the corresponding SD (FRG) equations and try
to solve  them, given that the gap equation \eqref{d22} must be
solved simultaneously. We expect that  different kinds of
solutions, similar to those found in \cite{Fischer:2008uz}, will emerge. A
way to distinguish between them could be based on selecting the most stable
solution, i.e. the one with the lowest corresponding vacuum energy. We
notice that there is still a lot of information available about the action %
\eqref{snul}, e.g. nonrenormalization properties typical  of the
Landau gauge, a  renormalizable softly broken Slavnov-Taylor
identity, etc. \cite{Dudal:2008sp}. We conclude that in the current spirit
of using the KO condition \eqref{KO0} as in \cite%
{Fischer:2008uz,Sternbeck:2008mv}, there is no clear connection
between the KO criterion $u(0)=-1$ and the highly nontrivial issue of
confinement. All that  one can say is that there is a violation
of positivity in the gluon propagator, which is indicative of confinement,
but certainly not a proof of it.  Also, in the light of our
previous results, we disagree with the statement made in \cite%
{Maas:2008ri,Fischer:2008uz,Sternbeck:2008mv} about the fact that the
SD(FRG) solution with an infrared enhanced ghost propagator would refer to
the absolute Landau gauge, i.e. to  the restriction to the FMR $%
\Lambda $.  Unfortunately, at present, a way to implement
the restriction to the FMR $\Lambda $ remains completely unknown. Moreover,
we remind that the recent lattice data have given quite clear evidence about
the fact that the ghost propagator is not enhanced in the infrared, within
the current accuracy of implementing the Landau gauge as the minimum of the
functional $\int d{^{d}}x(A_{\mu }^{u})^{2}$, $u\in SU(N)$ \cite%
{Cucchieri:2007md}. Even if in the future more powerful algorithms would
bring the simulations closer to the FMR $\Lambda $, the ghost propagator
will not get more enhanced than before, on the contrary \cite%
{Cucchieri:1997dx}. Moreover, we have shown in this letter that the KO
boundary condition is equivalent with the GZ framework, which explicitly
refers to  the restriction to the Gribov region $\Omega $. This
is irrespective of the fact that the ghost is enhanced or not, implementing
KO breaks the conventional BRST symmetry, and refers to $\Omega $, not to $\Lambda $. We wish to underline that implementing the boundary  as in eqs.\eqref{d21}
and \eqref{d22} will not necessarily give rise to an infrared enhanced ghost.
Additional nontrivial quantum effects can combine with the boundary effect,
we refer for instance to the effects of the operator $\overline{\varphi}%
_{\mu}^{ac}\varphi_{\mu}^{ac}-\overline{\omega}_{\mu}^{ac}\omega_{\mu}^{ac}$ in the case under study. We can make the analogy with spontaneous symmetry breaking: although the
starting action in that case enjoys a certain symmetry, nonperturbative
quantum effects can induce a shift from the symmetric, but unstable, vacuum\footnote{%
We refer to condensates which are dynamically favored by lowering the vacuum energy.}, causing qualitative and quantitative effects in the
theory.

In conclusion, we hope that this letter has clarified the
relation between the KO and GZ framework. Our main result is expressed by
eqs.\eqref{d21} and \eqref{d22}, which show that the KO and GZ frameworks
are equivalent, provided the KO boundary condition is properly taken into
account from the beginning. The conventional BRST operator \eqref{BRST}
suffers from a soft breaking, which relies precisely on the implementation
of the boundary condition.  Some ingredients in certain
formalisms, which we have tried to outline,  have thus to be
considered as assumptions rather than as proofs. In particular, the precise
relation between implementing the  KO criterion and confinement
remains to be clarified. One of the challenges lying ahead is how to define what the relevant physical operators are in the KOGZ
framework, if there is no (local) nilpotent BRST symmetry generator found.

\section*{Acknowledgments}
D.~Dudal and N.~Vandersickel are supported by the Research-Foundation
Flanders (FWO Vlaanderen). S.~P.~Sorella is supported by the FAPERJ, Funda{%
\c c}{\~a}o de Amparo {\`a} Pesquisa do Estado do Rio de Janeiro, under the
program \textit{Cientista do Nosso Estado}, E-26/100.615/2007. The Conselho
Nacional de Desenvolvimento Cient\'{\i}fico e Tecnol\'{o}gico (CNPq-Brazil),
the FAPERJ, the SR2-UERJ and the Coordena{\c{c}}{\~{a}}o de Aperfei{\c{c}}%
oamento de Pessoal de N{\'{\i}}vel Superior (CAPES) are gratefully
acknowledged for financial support.

\end{document}